\documentstyle[12pt]{article}
\newcommand{\half}{{1 \over 2}}
\newcommand{\beq}{\begin{equation}}
\newcommand{\eeq}{\end{equation}}
\newcommand{\bps}{\begin{psfrags}}
\newcommand{\eps}{\end{psfrags}}
\newcommand{\beqa}{\begin{eqnarray}}
\newcommand{\eeqa}{\end{eqnarray}}
\newcommand{\beqaa}{\begin{eqnarray*}}
\newcommand{\eeqaa}{\end{eqnarray*}}
\newcommand{\nonum}{\nonumber}
\newcommand{\ri}{\right}
\newcommand{\lf}{\left}

\newcommand{\Tr}{{\rm Tr}}

\newcommand{\del}{\partial}
\newcommand{\cob}{\delta}    %coboundary
\newcommand{\al}{\alpha}

\newcommand{\ep}{\epsilon}

\newcommand{\riya}{\rightarrow}

\newcommand{\bra}{\rangle}

\begin{document}

\sloppy
\sloppy
\sloppy
$\ $
%\begin{flushright}{ }
%\end{flushright}
\vskip 0.5 truecm
{\baselineskip=14pt
 \rightline{
 \vbox{
       \hbox{UT-777}
       \hbox{June  1997}
}}}
~~\vskip 5mm

\begin{center}
{\Large{\bf On a Lorentz covariant matrix regularization of\\ membrane
theories }}
\end{center}
\vskip .5 truecm
\centerline{\bf Kazuo Fujikawa and Kazumi Okuyama}
\vskip .4 truecm
\centerline {\it Department of Physics,University of Tokyo}
\centerline {\it Bunkyo-ku,Tokyo 113,Japan}
\vskip 0.5 truecm

\makeatletter
\@addtoreset{equation}{section}
\def\theequation{\thesection.\arabic{equation}}
\makeatother

\vskip 0.5 truecm

\begin{abstract}
A Lorentz covariant matrix regularization of membrane thories is studied.
It is shown that the action for a bosonic membrane can be defined by matrix
regularization in a Lorentz covariant manner. The generator of area
preserving diffeomorphism can also be
consistently defined by matrix regularization, and we can make the area 
preserving gauge symmetry manifest. However,  the reparametrization
BRST charge explicitly depends on a specific basis set 
introduced to define the matrix regularization. We also briefly comment on 
an extension of the present formulation to a supermembrane.

\par

\end{abstract}

\baselineskip 6mm
%-----------------------------------------
\section{Introduction}
%-----------------------------------------
The dynamics of quantum membranes is not well understood yet[1] - [14].
First of all, the world volume dynamics of membranes is not renormalizable
not only for 
bosonic membranes but also for supermembranes. As an attempt to understand
the dynamics of membranes better, a matrix regularization[8][12] received
much attention recently [15] - [18].  It is hoped that the matrix
regularization, if
properly treated ,might 
lead to a non-perturbative treatment of quantized membranes. 

The matrix regularization of a supermembrane in Ref.[8] is based on the
light-cone gauge formulation. In fact, the covariant quantization of the
supermembrane is  involved , and 
 for this reason we here mainly study a Lorentz covariant matrix
regularization of bosonic membranes. A brief comment on a supermembrane is
made at the 
end of this note. It is true that certain crucial aspects of quantized
membranes can only  be studied by the supermembrane, but all may not be
lost by considering  bosonic membranes,  and some of important dynamical
issues of the quantum theory of membranes may still be studied by examining
the detailed dynamics of bosonic membranes. It is our hope that a Lorentz
covariant matrix regularization  of bosonic membranes may lead to a better
understanding of quantum theory of membranes in general. 

The starting Lagrangian of the bosonic membrane is given by  
\begin{eqnarray}
 {\cal L}_{0} &=& \half \sqrt{-g} ( 1-g^{ab}\del_aX^{\mu}  \del_bX_{\mu})
\nonum \\
  &=& -\half \det \tilde{g}^{ab} - \half\tilde{g}^{ab}\del_aX^{\mu} 
\del_bX_{\mu}
\end{eqnarray}
where we defined 
\begin{equation}
 \tilde{g}^{ab}= \sqrt{-g}g^{ab}
\end{equation}
and the indices $a$ and $b$ run over $0$ to $2$. If one generalizes  the
conformal gauge in string theory to the case of the present membrane
theory, one has 
\begin{equation}
 g^{00}+ \det g^{kl}=0\ \  ,and \ \  g^{0k}=0    
\end{equation}
where $k, l = 1,2$, or in an  equivalent notation
\begin{equation}
\tilde{g}^{00}+1=0\ \   ,and \ \  \tilde{g}^{0k}=0 
\label{our-gauge}
\end{equation}
In this gauge, the world volume of membranes is reduced to a product of
two-
dimensional space $\Sigma$ times time coordinate $R$, $\Sigma \times R$.
The Faddeev-Popov gauge fixing Lagrangian is given by  
\begin{eqnarray}
 {\cal L}_g &=& N_a (\tilde{g}^{0a}+ \cob^{0a})\nonumber \\
  &+& i b_a \lf[\del_b (c^b \tilde{g}^{0a})- \tilde{g}^{ba}\del_b c^0
- \tilde{g}^{0b}\del_b c^a\ri]
\end{eqnarray}
where $N_{a}$ are  the Lagrangian multiplier fields , and $c^{a}$ and
$b_{a}$ stand for ghosts and 
anti-ghosts, respectively. The total Lagrangian 
\begin{equation}
{\cal L} = {\cal L}_{0} + {\cal L}_{g}
\end{equation}
is written as[9] 
\begin{equation}
 {\cal L} = \half \del_0X^{\mu}  \del_0X_{\mu}- \half \det G_{kl}
+i b_0 (\del_0 c^0 -\del_k c^k) +i b_k \del_0 c^k 
\end{equation}
with
\begin{equation}
 G_{kl}=\del_kX^{\mu}  \del_lX_{\mu} +ib_k \del_l c^0 +ib_l \del_k c^0
\end{equation}
after integration over $N_a$ and $ \tilde{g}^{kl}$.
The Lorentz covariant Lagrangian has a structure quite similar to that of
the light-cone Lagrangian, though it contains up to quartic couplings of
ghost fields.

%-------------------------------------------
\section{Symplectic Structure}
%-------------------------------------------
To implement  the matrix regularization, it is important to understand the 
symplectic structure. For this purpose, it is convenient to use the form
notation. The symplectic form on the  2d manifold $\Sigma$, whose canonical
coordinates are $\sigma^1$ and $\sigma^2$, is defined by
\begin{equation}
 \omega = d \sigma^1 \wedge  d \sigma^2
\end{equation}
The Hamiltonian vector field $\vec{f}$ for  a function
$f(\sigma^1,\sigma^2)$
is defined  by 
\begin{equation}
  \vec{f} = \del_1 f \del_2 -\del_2 f \del_1 
\end{equation}
The ``Poisson bracket'', when  we regard the coordinates $\sigma^1$and
$\sigma^2$ as two-dimensional phase space variables, is written in a number
of  ways 
\begin{equation}
 \{f,g\}\equiv {\cal{L}}_{\vec{f}}g = \omega
(\vec{f},\vec{g})=(-df,\vec{g})
=\del_1 f \del_2 g -\del_2 f \del_1 g
\label{p-bracket}
\end{equation}
where ${\cal{L}}_{\vec{f}}$ stands for the Lie derivative generated by the 
Hamiltonian vector field $\vec{f}$,\ \ $\omega (\vec{f},\vec{g})$ is the
inner product of two Hamiltonian  vector fields $\vec{f}$ and $\vec{g}$
with the two form $\omega$, and  the notation $( \  ,\  )$ stands for a
contraction of a 1-form and a vector field. 
The  Poisson  bracket in (\ref{p-bracket}) satisfies 
\begin{eqnarray}
\int d^2 \sigma \{f,g\}h &=&\int d^2 \sigma f\{g,h\}\nonumber\\
df \wedge dg &=& \{f,g\}\omega 
\end{eqnarray}
 In our analysis, it is convenient to regard the 
space components of $b_a,c^a$  as 1-form and vector field on $\Sigma$,
respectively
\begin{eqnarray}
 \bf{b} &=& b_1 d \sigma^1 + b_2 d \sigma^2 \nonumber\\
 \bf{c} &=& c^1 \del_1 + c^2 \del_2
\end{eqnarray}
 Our Lorentz covariant Lagrangian is then written as 
\begin{equation}
{\cal L}=\half \del_0X^{\mu}  \del_0X_{\mu}-\half \det G_{kl}
+i b_0 (\del_0 c^0 -{\rm{div}} {\bf{c}} ) + i ({\bf{b}}, \del_{0} 
{\bf{c}})
\label{cov-lag}
\end{equation}
with
\begin{equation}
\det G_{kl}=\sum_{\mu < \nu}\{X^{\mu},X^{\nu}\}\{X_{\mu},X_{\nu}\}
+2i ({\bf{b}},\vec{X_{\mu}})\{X^{\mu},c^0\}
-3({\bf{b}},\vec{c^0})^2
\end{equation}
where $\vec{X}_{\mu}$ and $\vec{c}^{0}$ are Hamiltonian vector fields
associated with $X_{\mu}$ and $c^{0}$. 
The Lie derivatives of vector field $v$ and 1-form $\alpha$ are
respectively given by $
({\cal L}_u v)^k =  u^l \del_l v^k - \del_l u^k v^l $ and $
({\cal L}_u \al)_k =  u^l\del_l \al_k + \del_k u^l\al_l $.

If one uses a gauge $\tilde{g}^{00} = - \rho (\sigma_{1},\sigma_{2})$
instead of  (\ref{our-gauge}), the Lagrangian (1.7) is written as 
\begin{equation}
\rho^{-1} {\cal L} = \half \del_0X^{\mu}  \del_0X_{\mu}- \half \det G_{kl}
+i b_0 (\del_0 c^0 -\rho^{-1}\del_k (\rho c^k)) +i b_k \del_0 c^k 
\end{equation}
and the symplectic form becomes $\omega = \rho  d \sigma^1 \wedge  d
\sigma^2$. The Poisson bracket (\ref{p-bracket}) is then replaced by 
\begin{equation}
 \{f,g\} = \rho^{-1}( \del_1 f \del_2 g -\del_2 f \del_1 g )
\end{equation}
Also, ${\rm{div}} {\bf{c}}= \rho^{-1}\del_k (\rho c^k)$  and $   V 
=\rho^{-1} (\del_1 b_2 - \del_2 b_1) = \ast d{\bf{b}}$ in (3.9) below. The 
$\rho$ plays a role of density [8]. In this paper  we work with the gauge
(1.4).
%-------------------------------------------
\section{Canonical Formalism and Symmetry Properties}
%-------------------------------------------
 
The Hamiltonian corresponding to our covariant Lagrangian is given by
\begin{equation}
{\cal{H}}=\half P^{\mu}P_{\mu}+\half \det G +ib_0\del_k c^k
\label{hamiltonian}
\end{equation}
with the basic canonical commutation relations
\begin{eqnarray}
[X_{\mu}(\sigma),P_{\nu}(\sigma ')] &=& 
        i\eta_{\mu\nu}\cob^{(2)} (\sigma-\sigma ') \\
\lf[ b_a (\sigma),c^b (\sigma ')\ri]_{+} &=& 
        \cob^b_a \cob ^{(2)}(\sigma-\sigma ')
\end{eqnarray}
where $P_{\mu}(\sigma) = \partial_{0}X_{\mu}(\sigma)$ , and  $[ \ , \ ]_+$
stands
for an anti-commutator of operators.

We now  analyze the equations of motion. We start with the equations of
motion for $c^{0}$ and $b_{k}$  
\begin{eqnarray}
\del_0 c^0 -{\rm{div}}{\bf{c}} &=& 0\nonumber \\
\del_0 b_k -\del_k b_0   &=& 0
\end{eqnarray}
where $k= 1,2$,  and the 1st equation corresponds to the conservation of
``current'' $c^a$,
and the  2nd equation shows that the  $0k-$components of $db_a$ ( in a
3-dimensional notation) vanishes.
The  equations of motion for ${\bf{c}}$ and $b_0$ are given by 
\begin{eqnarray}
\del_0 {\bf{c}} &=& \vec{X_{\mu}}\{X^{\mu},c^0 \}
      + 3i\vec{c^0}({\bf{b}},\vec{c^0}) \nonumber \\
\del_0 b_0 &=& -\{({\bf{b}},\vec{X_{\mu}}),X^{\mu}\}
     -3i \ast\lf({\cal{L}}_{\vec{c^0}}{\bf{b}} \wedge {\bf{b}} \ri)
\end{eqnarray}
where the star ($\star$) operation transforms a two form to its dual (i.e.,
scalar in the present case). 
The constant shift in  $b_0$, which is a symmetry of our action,  is
generated by 
\begin{eqnarray}
 b_0 &\riya& b_0 + \xi \\
{\rm{charge}} &=& \int d^2 \sigma c^0 
\end{eqnarray}
Our Lagrangian is also invariant under an  addition of a (time independent)
hamiltonian vector field $\vec{w}$ to ${\bf{c}}$, and this symmetry  is 
generated by the operator $V$ defined by       
\begin{eqnarray}
 {\bf{c}}  &\riya&  {\bf{c}}+ \vec{w}\\
      V  &=& \del_1 b_2 - \del_2 b_1 = \ast d{\bf{b}}
\label{def-V} 
\end{eqnarray}
In fact,  the ghost part of the action (\ref{cov-lag}) is invariant under $
 {\bf{c}}
 \riya  {\bf{c}}+ {\bf{f}}$ with a more general(time independent) vector
field ${\bf{f}}$
with
$\rm{div}{\bf{f}} =0$. 
The basic BRST transformation  in our covariant formulation  is defined by
\begin{eqnarray}
\cob_{BRST} X^{\mu} &=& \ep c^a \del_a X^{\mu}  \nonum \\
\cob_{BRST} c^a &=& \ep c^b \del_b c^a  \nonum \\
\cob_{BRST} b_a &=& i\ep B_a 
\end{eqnarray}
where $\epsilon$ is a Grassmann parameter, and 
\begin{equation}
 B_0=\half (\del_0 X^{\mu})^2+ \half \det G +i\del_0 b_0 c^0
+i\del_k b_0 c^k+ 2i b_0 \del_0 c^0 +i b_k \del_0 c^k 
\end{equation}
\begin{equation}
 B_k= \del_0 X^{\mu} \del_k X_{\mu}  +i\del_0 (b_k c^0)
+i\del_l b_k c^l+ i b_l \del_k c^l + 2i b_0 \del_k c^0 
\end{equation}
These variables $B_{a}$, up tp equations of motion,  correspond to the
components $T_{0a}$ of the energy momentum tensor  on the world volume. 
The BRST charge is defined by[9] 
\begin{eqnarray}
Q &=& \int d^2 \sigma \Biggl[ c^0\Bigl(\half (\del_0 X^{\mu})^2+ \half \det
G\Bigr) 
\Biggr. \nonum \\
&&  \Biggl.+c^k \del_0 X^{\mu} \del_k X_{\mu} 
-ib_0(c^0 \del_k c^k+c^k \del_k c^0)-ib_k c^l \del_l c^k  \Biggr]
\label{brst-charge} 
\end{eqnarray}
By using the BRST charge and the operator $V$ , the generator $L$ of the
area preserving diffeomorphism  is given by[13]
\begin{equation}
L=[Q,V]_+ = \del_1 B_2 - \del_2 B_1 =\ast d{\bf{B}}
\label{def-L} 
\end{equation}
 and
\begin{equation}
L_w = \int d^2 \sigma wL =\int d^2 \sigma (- {\bf{B}},\vec{w})
\end{equation}
We here defined   
\begin{eqnarray}
{\bf{B}}&=& \del_0 X^{\mu} dX_{\mu}+i\del_0({\bf{b}}c^0)
             +2ib_0 dc^0 -i{\cal{L}}_{{\bf{c}}}{\bf{b}}\nonumber \\
L&=& \{\del_0 X^{\mu} ,X_{\mu}\}+ 2i\{b_0,c^0 \}
+i\del_0({\bf{b}},\vec{c^0})
     +i\del_0 (Vc^0) +iV c+i(dV,{\bf{c}})
\label{form-of-L}
\end{eqnarray}
with  $c={\rm{div}}{\bf{c}}$. If we use the equations of motion for ghost
variables, we can rewrite $\bf{B}$ and $L$ as  
\begin{eqnarray}
{\bf{B}}&=& \del_0 X^{\mu} dX_{\mu}+ib_0 dc^0 
             +id(b_0 c^0) +i{\bf{b}}c
-i{\cal{L}}_{{\bf{c}}}{\bf{b}}\nonumber
 \\
L&=& \{\del_0 X^{\mu} ,X_{\mu}\}+ i\{b_0,c^0 \} +i({\bf{b}},\vec{c})
     +2iV c+i(dV,{\bf{c}})
\end{eqnarray}

 The transformation laws of area preserving diffeomorphism are  given as
\begin{eqnarray}
-i[L_w,X ]&=& \{w,X\}  \nonum \\
-i[L_w,c^0 ]&=& \{w,c^0\}  \nonum \\
-i[L_w,b^0 ]&=& \{w,b^0\}  \nonum \\
-i[L_w,c^1 ]&=& \{w,c^1\}-{\cal{L}}_{{\bf{c}}}(\vec{w})^1
                       = ({\cal{L}}_{\vec{w}}{\bf{c}})^1  \nonum \\
-i[L_w,c^2]&=& \{w,c^2\}-{\cal{L}}_{{\bf{c}}}(\vec{w})^2
                       = ({\cal{L}}_{\vec{w}}{\bf{c}})^2  \nonum \\
-i[L_w,b_1]&=& \{w,b_1\}+({\bf{b}},\del_1 \vec{w})
                       =({\cal{L}}_{\vec{w}}{\bf{b}})_1  \nonum \\
-i[L_w,b_2]&=& \{w,b_2\}+({\bf{b}},\del_2 \vec{w})
                       =({\cal{L}}_{\vec{w}}{\bf{b}})_2  
\end{eqnarray}
In these equations, $[ \  ,\ ]$ in the left-hand side is a commutator of
operators and $\{ \ ,\  \}$ in the right-hand side is a Poisson bracket.
In general, the transformation law of an operator ${\cal O}$ is written as 

\begin{equation}
i[L_w,{\cal{O}} ]=  -{\cal{L}}_{\vec{w}}{\cal{O}}
\end{equation}
with $\vec{w}$ a Hamiltonian vector field associated with $w$.

We here note interesting  algebraic relations  satisfied by the symmetry
generators
 $V_{\xi}=\int d^2 \sigma \xi V$ and $L_w=\int d^2   \sigma w L$ in our 
Lagrangian
\begin{eqnarray}
[V_{\xi},V_{\eta}] &=&  0   \nonum  \\
\lf[ L_{w},V_{ \xi } \ri]  &=& iV_{ \{w,\xi \}}  \nonum \\
\lf[ L_{f},L_{g} \ri]  &=& iL_{ \{f,g\} }
\end{eqnarray}
where $V$ is defined in (\ref{def-V}). 

In connection with the area preserving gauge symmetry, we note that  the
action may be defined anew by using the Hamiltonian in (\ref{hamiltonian})
as
\begin{equation}
{\cal L}= P^{\mu}\del_0 X_{\mu}+ib_a \del_0 c^a -{\cal{H}} +\cob_{BRST}
(iAV)
\label{modified-lag}
\end{equation}
where we added a BRST exact term to the action , which does not change the 
physical sector of the Fock space, 
\begin{eqnarray}
\int d^{2}\sigma \cob_{BRST} (iAV) &=& \int d^{2}\sigma ( i\lambda V-AL)  
\nonum \\
&=& \int d^{2}\sigma [i({\bf{b}},\vec{\lambda})+({\bf{B}},\vec{A})]
\label{brst-exact-term}
\end{eqnarray}
The transformation properties of the auxiliary fields $A$ and $\lambda$
under 
the BRST transformation, 
the symmetry genrated by $V$ and the area preserving diffeomorphism
generated 
by $L$ are respectively defined by
\begin{equation}
\begin{array}{lll}
\cob_{BRST} A = \lambda,  &  \cob_{BRST} \lambda = 0& \\
\cob_V {\bf{c}} = \vec{\xi},  &  \cob_V \lambda = D_0 \xi & \\
\cob_L A =  D_0 w,  &  \cob_L \lambda = -\{w,\lambda\}&\\
 \end{array}
\end{equation}
and for other generic variables 
\begin{equation}
\cob_L{\cal{O}} = -{\cal{L}}_{\vec{w}}{\cal{O}}
\end{equation}
where we defined $D_0 f = \del_0 f+\{A,f\}$. The significance of two
auxiliary 
variables $A$ and $\lambda$ becomes transparent when one integrates over 
$P_{\mu}$ in the above Lagrangian (3.21). We then obtain 
\begin{equation}
{\cal L}=\half (D_{0}X^{\mu})^{2} - \half \det G_{kl}
+i b_0 (D_{0}c^0 - {\rm{div}}{\bf{c}} ) + i ({\bf{ b}}, D_{0}{\bf{c}}
+\vec{\lambda})  
\label{gauged-lag}
\end{equation}
where  we defined 
\begin{equation}
D_{0}{\cal O} = \partial_{0}{\cal O} + {\cal L}_{\vec{A}}{\cal O}
\end{equation}
for a general operator ${\cal O}$. By this way, we can make the area
preserving gauge symmetry explicit in (3.25), and the auxiliary variable
$A$ corresponds to the 
gauge field for this symmetry and $\lambda$ is a BRST partner of $A$; 
$\lambda$ can be regarded as a gauge field for the symmetry generated by 
$V$, as is expected from the first expression in (3.22).   In this
formulation, our original covariant 
Lagrangian corresponds to the gauge fixing $A=\lambda = 0$ of area
preserving gauge 
symmetry:This is realized by adding the gauge fixing terms $NA - i\xi
\lambda$
to (3.25) with a Nakanishi-Lautrup doublet $(\xi, N)$.
See also Ref.[8] for a related analysis in the light-cone gauge.
In the following we work with our original covariant Lagrnagian
(\ref{cov-lag}). 

%-------------------------------------------
\section{Solving Constraints and Matrix Regularization}
%-------------------------------------------
We have  constraints on the physical states in our formulation 
\begin{eqnarray}
&& V|phys \bra =  d{\bf{b}} |phys \bra =0  \label{V=0} \\
&& L|phys \bra =  d{\bf{B}} |phys \bra =0  \label{L=0}
\end{eqnarray}
in addition to the BRST symmetry $Q|phys \bra = 0$. 
We solve the first constraint (\ref{V=0}) in the operator level by writing 
{\bf{b}} in a {\em locally} exact form
\begin{equation}
{\bf{b}}=-db
\end{equation}
and thus $V\simeq 0$ in the sense of Dirac in (\ref{def-L}). At the same
time, we
introduce a new variable defined by
\begin{equation}
c\equiv \rm{div} \bf{c}
\label{c=divc}
\end{equation}
Note that $c$ is invariant under the transformation generated by $V$.
The operations in (4.3) and (4.4) are regarded as a gauge fixing of 
$V$-symmetry by using the gauge condition  $F=\del_1 c^2-\del_2 c^1 =0$.
This is seen 
by the following path integral manipulation
\begin{eqnarray}
&&\int{\cal D}{\bf{b}}{\cal D} {\bf{c}}\cob(V)\cob(F)
\det\{V,F\}^{-1}\exp(iS) \nonum \\
&=&\int {\cal D}{\bf{b}} {\cal D}{\bf{c}}\cob(V)\cob(F)
\det\triangle^{-1}\exp(iS) \nonum \\
&=& \int {\cal D}b {\cal D}c\exp(iS')
\end{eqnarray}
The Lagrangian (\ref{cov-lag}) is then written as 
\begin{equation}
{\cal L}'=\half \del_0X^{\mu}  \del_0X_{\mu}-\half \det G'
+i b_0 (\del_0 c^0 -c ) + ib\del_{0}c
\label{V=0lag}
\end{equation}
with
\begin{equation}
\det G'=\sum_{\mu < \nu}\{X^{\mu},X^{\nu}\}\{X_{\mu},X_{\nu}\}
+2i \{b,X_{\mu}\}\{X^{\mu},c^0\}
-3\{b,c^0 \}^2
\end{equation}
All the  variables here are treated  as functions on $\Sigma$, and the
variables $b$ and $c$ become  canonical conjugate to each other.
If we use the equations of motion for  ghost variables 
\begin{eqnarray}
\del_0 c^0 -c &=& 0 \nonumber\\
\del_0 b +b_0   &=& 0
\end{eqnarray}
 the generator of the area preserving diffeomorphism in (3.17) is further
rewritten as  
\begin{eqnarray}
L|_{{\bf{b}}=-db}&=& \{\del_0 X^{\mu},X_{\mu}\}
                      +2i \{b_0,c^0\}+ i\del_0 \{b,c^0\}  \nonum \\
  &=& \{\del_0 X^{\mu},X_{\mu}\}+i \{b_0,c^0\}+i \{b,c\}\equiv L^{\prime}
\end{eqnarray}
The transformation law of general variable  ${\cal O}$ appearing in (4.6) 
under the area preserving diffeomorphism  is  written as 
\begin{equation}
\cob {\cal{O}}=\{w,{\cal{O}}\} 
\end{equation}
with the generator of   this transformation being  given by $L^{\prime}$
(4.9).

We solved the constraint $V=0$ in the operator level, but its BRST
transform
$L^{\prime} = 0$ is not solved in the operator level in our treatment. The 
BRST symmetry, which is manifest in (\ref{V=0}) and (\ref{L=0}), is no more
manifest
after
solving $V=0$.
The gauge symmetry generated by $V$ (and also by $L$) is characterized by a
time independent 
parameter $\vec{w}$, and in this sense it is non-dynamical and analogous to
the residual symmetry of  $A_{0}=0$ gauge for Yang-Mills fields. This
symmetry 
plays an important role to eliminate non-oscillatory (instability) modes
when
one compactifies two spatial coordinates in membrane theory[13].  

In passing, the ``Poisson 
bracket'' of two fermionic( Grassmann) variables $\xi$ and $\eta$ are
defined by
\begin{equation}
\{\xi ,\eta \} \equiv \partial_{1}\xi\partial_{2}\eta -
\partial_{2}\xi\partial_{1}\eta  
\end{equation}
One can then confirm 
\begin{equation}
\{\xi ,\eta\} = \{\eta ,\xi\}
\end{equation}
by noting the anti-commuting property of $\xi$ and $\eta$. In the matrix
regularization, the fermionic variables are then specified  by the {\em
anti-commutator} of two
matrix valued fermionic variables.

The Lagrangian (4.6)   and the generator of the area preserving
diffeomorphism (4.9)  are both written solely in terms of  Poisson
brackets or bi-linear combinations of two field variables.  It is thus
straightforward 
to introduce the matrix regularization by formally replacing the Poisson
bracket by 
an (anti-)commutator of matrix valued operators. This transition is
facilitated by 
expanding the various operators in terms of a complete set of basis
vectors defined on the two-dimensional space  $\Sigma$ .

For example,
\begin{eqnarray}
X^{\mu}(\tau, \sigma_{1}, \sigma_{2})& \equiv& \sum_{A} X_{A}^{\mu}(\tau)
Y^{A}(\sigma_{1}, \sigma_{2})\nonumber\\
b(\tau, \sigma_{1}, \sigma_{2})& \equiv& \sum_{A} b_{A}(\tau)
Y^{A}(\sigma_{1}, \sigma_{2})
\end{eqnarray}
and similarly for other variables with a suitable complete orthonormal set
of functions $\{ Y^{A}(\sigma_{1}, \sigma_{2})\}$ on the space $\Sigma$. We
then formally replace the set $\{ Y^{A}(\sigma_{1}, \sigma_{2})\}$ by the
(hermitian)
generators of the group $SU(N)$ with $N\rightarrow \infty$ [8]. The
dynamical 
varibles  are then promoted to (infinite dimensional)  matrices 
and we have , for example,
\begin{eqnarray}
\int d^{2}\sigma \partial_{0}X^{\mu}\partial_{0}X_{\mu}& = & \Tr
\partial_{0}X^{\mu}\partial_{0}X_{\mu} 
\nonumber\\
\int d^{2}\sigma \det G'&=& \Tr\Bigl( - \sum_{\mu <
\nu}[X^{\mu},X^{\nu}][X_{\mu},X_{\nu}] \Bigr.\nonum \\
&&\Bigl.- 2i [b,X_{\mu}][X^{\mu},c^0]
+ 3[b,c^0 ]_{+}[b,c^0 ]_{+} \Bigr)
\label{det-in-matrix}
\end{eqnarray}
where $\Tr$ specifies the trace operation, and  $[\ \ , \ \ ]$ stands for 
the commutator and $[\ \ , \ \ ]_{+}$ for the 
anti-commutator.
In the matrix notation, we can write the action, the Hamiltonian and the
generator of the area
preserving diffeomorphism respectively as 
\begin{eqnarray}
\int d\tau d^{2}\sigma {\cal L} &=&
\int d\tau \Tr \Bigl( \half \del_0X^{\mu}  \del_0X_{\mu}-\half \det G'
+i b_0 (\del_0 c^0 -c ) + ib\del_{0}c \Bigr)\nonumber\\
H &=& \int d^{2}\sigma {\cal H} = \Tr \Bigl( \half P^{\mu}P_{\mu} 
+ \half \det G^{\prime} + ib_{0}c \Bigr)\nonumber \\
\int d^{2}\sigma w(\sigma) L^{\prime} &=&
\Tr w \Bigl( -i[\del_0 X^{\mu},X_{\mu}] + [b_0,c^0]_{+} +[b,c]_{+} \Bigr) 
\label{final-matrix-lag}
\end{eqnarray}
where $P^{\mu} = \partial_{0}X^{\mu}$  and we make the replacement 
 (4.14)
inside $\det G^{\prime}$.  The parameter $w(\sigma)$ is also expanded in
terms of the basis set
$\{ Y^{A}(\sigma_{1}, \sigma_{2})\}$.

%---------------------
{\section{Discussion} 
%--------------------
We have shown that the action and the generator of area preserving
diffeomorphism for bosonic membranes can be consistently defined by Lorentz
covariant matrix regularization, if one keeps $N$ of $SU(N)$ in (4.15)
finite. Our matrix regularized Lagrangian has a structure very similar to
that of the light-cone gauge[8]. This suggests that the basic dynamics of
membrane theory is not much influenced by the infinite 
momentum frame which is the underlying physical picture of the light-cone
gauge. However, it is important to recognize that the BRST charge
(\ref{brst-charge}) in
our formulation 
cannot be written in a simple matrix notation by using commutators for 
bosonic variables and anti-commutators for fermionic variables even in the
formal way.
For example, we cannot make the replacement in (\ref{c=divc}) consistently,
and the
variables $c^{k}$ survive in the BRST charge till the end. This means that
one cannot
eliminate the 
explicit dependence on a particular basis set $\{ Y^{A}(\sigma_{1},
\sigma_{2})\}$ in the Lorentz covariant formulation, while   the basis set
$\{ Y^{A}(\sigma_{1}, \sigma_{2})\}$ carries the information of the
topology of world volume. 

Our analysis suggests that Lorentz covariance and unitarity ( or BRST 
symmetry) , both of which are essential for any sensible theory, cannot
be simultaneously maintained in matrix regularization in a   manner
independent of  world volume topology.  Although it is premature to make
a definite statement at this moment, our 
consideration at least appears to be consistent with the recent analysis by
Matsuo et al.\cite{matsuo}. They examined the Lorentz covarince of membrane
theory in
 the (manifestly unitary) light-cone gauge formulation, and they conclude
that the 
explicit appearance of the world volume metric , which contains the
information
of  topology, is required to recover the full Lorentz covariance.   

Ultimately, one may want to analyze the dynamics of supermembrane in a
Lorentz
covariant manner not only in continuum notation but also in matrix
regularization. The $32$-component spinor $\theta
(\tau,\sigma_{1},\sigma_{2})$ in $d=11$ dimensional space is irreducible,
and any {\em algebraic} gauge 
fixing of the $\kappa$ -symmetry associated with $\theta
(\tau,\sigma_{1},\sigma_{2})$ generally breaks the Lorentz symmetry
$SO(10,1)$. One possible algebraic gauge fixing of $\kappa$ symmetry is to
decompose 
\begin{equation}
\theta (\tau,\sigma_{1},\sigma_{2}) = \theta_{L} + \theta_{R}
\end{equation}
 and to impose the gauge condition $\theta_{R}  = 0$\cite{kall,berg,APS}.
This breaks 
the symmetry $SO(10,1)$ down to $SO(9,1)$. If one combines this algebraic 
gauge fixing with the $SO(10,1)$ covariant treatment of reparametrization
symmetry in Ref.[9] one can define an $SO(9,1)$ invariant supermembrane
theory consisting of polynomials up to quartic terms. The theory thus
defined  is located in between the full
covariant treatment and the light-cone gauge. This formulation may be
useful 
in the analysis of dynamical issues related to
M-theory\cite{duff,bfss,townsend}, for example.
The supersymmetry algebra in this formulation exhibits interesting
properties, and a detailed account of this analysis will be reported
elsewhere.

\end{document}